\documentclass[10pt, a4paper, twocolumn, twoside]{article} 
\usepackage{amsfonts, amsmath, amsthm, amssymb} 

\usepackage[english]{babel} 

\usepackage{microtype} 

\usepackage{amsmath,amsfonts,amsthm} 

\usepackage[svgnames]{xcolor} 

\usepackage[small, labelfont=bf, up]{caption} 

\usepackage{booktabs} 

\usepackage{lastpage} 

\usepackage{graphicx} 

\usepackage{enumitem} 
\setlist{noitemsep} 

\usepackage{sectsty} 
\allsectionsfont{\usefont{OT1}{phv}{b}{n}} 

\usepackage{geometry} 

\usepackage[T1]{fontenc} 
\usepackage[utf8]{inputenc} 

\usepackage{XCharter} 

\usepackage{hyperref}

\usepackage{journals}

\usepackage{fancyhdr} 
\newcommand{\shorttitle}[1]{\fancyhead[CE]{\textsl{#1}}}
\newcommand{\shortauthors}[1]{\fancyhead[CO]{\textsl{#1}}}

\fancypagestyle{firstpage}{ 
	\fancyhf{}
        \lhead{\vspace*{0.0em} \textit{\footnotesize 22$^{\rm nd}$ European Workshop on White Dwarfs \\ S. Lauer, T. Rauch \\[-0.2em]
          August 15--19, 2022, T\"ubingen,  Germany}}
}

\date{}

\newcommand{\authorstyle}[1]{{\large\usefont{OT1}{phv}{b}{n}\color{DarkRed}#1}} 

\newcommand{\institution}[1]{{\footnotesize\usefont{OT1}{phv}{m}{sl}\color{Black}#1}} 

\usepackage{titling} 

\newcommand{\HorRule}{\color{DarkGoldenrod}\rule{\linewidth}{1pt}} 

\pretitle{
	\vspace{-30pt} 
	\HorRule\vspace{10pt} 
	\fontsize{16}{12}\usefont{OT1}{phv}{b}{n}\selectfont 
	\color{DarkRed} 
}

\posttitle{\par\vskip 10pt} 

\preauthor{} 

\postauthor{ 
	\vspace{0pt} 
	{\par\HorRule} 
	\vspace{-10pt} 
}

\newcommand{\newabstract}[1]{
    {\section*{Abstract}
    \bfseries #1}
  }

\usepackage[authoryear]{natbib}
\title{Uncertainties in the 12C+12C reaction rate and their impact on the composition of ultra-massive WDs} 

\shorttitle{Uncertainties in the carbon burning nuclear reactions} 

\shortauthors{De Gerónimo, Miller Bertolami, Catelan, Battich} 

\author{
  \authorstyle{F.~C.~De Gerónimo,$^{1,2}$ M.~M.~Miller~Bertolami,$^3$, M.~Catelan$^{1,2}$ and T.~Battich$^4$}
	\newline\newline 
 $^1$\institution{Instituto de Astrof\'{\i}sica, Pontificia Universidad Cat\'olica de Chile}\\
$^2$\institution{Millennium Institute of Astrophysics}
            \\
	$^3$\institution{Facultad de Ciencias Astron\'omicas y Geof\'isicas, Universidad
          Nacional de La Plata, Argentina}\\ 
	$^4$\institution{Max-Planck-Institut für Astrophysics, Garching bei München, Germany}\\ 
    }  

\begin{document}

\maketitle 

\thispagestyle{firstpage} 


\newabstract{Stars with initial masses  $7 \, M_{\odot} \lesssim M_{\rm ZAMS} \lesssim 9 \, M_{\odot}$ reach temperatures high enough to ignite C under degenerate conditions after the end of He-core burning \citep{1994ApJ...434..306G}. These isolated stars are expected to evolve into the  so-called super AGB (SAGB) phase and may end their lives as ultra-massive ONe WDs \citep[see][and references therein]{2006A&A...448..717S, 2007A&A...476..893S, 2010A&A...512A..10S, 2019A&A...625A..87C}. 
  The exact proportions of O and Ne found in the core at
  the end of the SAGB phase will determine the cooling times and
  pulsational properties of these WDs. Uncertainties affecting the rates of nuclear reactions occurring during the C burning phase should have a measurable impact on the distribution of $^{16}$O, $^{20}$Ne, $^{23}$Na and $^{24}$Mg and, consequently, on the evolution of the WD.
  
    Here we present a study of the impact of uncertainties in the
  $^{12}$C$(^{12}$C$,\alpha)^{20}$Ne and $^{12}$C$(^{12}$C$,p)^{23}$Na nuclear  reaction rates (and their branching ratios) on the chemical structure  of intermediate- to high-mass progenitors at the end of the
  C-burning phase. Using the stellar evolution code Modules for
  Experiments in Stellar Astrophysics ({\tt MESA}) we computed evolutionary sequences for stars with   initial masses 7.25$\leq M_{ZAMS}/M_{\odot}\leq$ 8.25, from the ZAMS to the SAGB   phase, adopting different prescriptions   for the $^{12}$C+$^{12}$C burning rates. 
  We found that adopting  lower reaction rates for the  $^{12}$C+$^{12}$C burning delays C-ignition by at most 2700~yrs, and the  ignition takes place in a position further from the center. 
  Our results shows that differences in   the $^{20}$Ne central abundances remain modest, below 14\%. 
}


\section*{Methodology \& Input physics}

We computed the complete evolution of progenitors stars of 7.25$\leq M_{\rm ZAMS}/M_{\odot}\leq$ 8.25 from the zero age main sequence (ZAMS) to the beginning of the SAGB phase.
During the carbon burning phase, 
we adopted the total nuclear reaction rate values from  \cite{1988ADNDT..40..283C}  (CF88) and the recently provided by \cite{2022A&A...660A..47M} Be test both the model based on the fusion-hindrance phenomenon (HIN rate) and the fusion-hindrance plus resonance model (HINRES), see Fig. \ref{fig:rates}. Both CF88 and HINRES rates shows very similar behaviour in the temperature range of C-burning.
Branching ratios of 65\% for the alpha exit channel, $^{12}$C$(^{12}$C$,\alpha)^{20}$Ne, and 35\% for the proton exit channel,  $^{12}$C$(^{12}$C$,p)^{23}$Na, were adopted according to \cite{2013ApJ...762...31P}. 
Additional evolutionary sequences with branching ratios 56-44 \citep{1988ADNDT..40..283C} were also computed. For a better comparison, overshooting was adopted only during the central H- and He-burning phases.

\begin{center}
\includegraphics[width=1.\columnwidth]{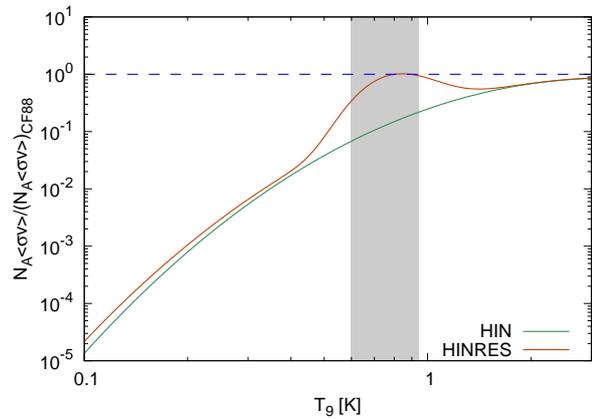}
\captionof{figure}{Normalized reaction rates from \cite{2022A&A...660A..47M}. Shaded region represent the temperature region of C-burning. }
\label{fig:rates}
\end{center}
%







\section*{Results}

C-ignition occurs for $M_{ZAMS}\geq7.25 M_{\odot}$ for CF88 and HINRES rates, and for  $M_{ZAMS}\geq7.30 M_{\odot}$ for the HIN rate. 
The ignition is delayed  by at most 2700 yrs for lower reaction rate and it occurs at a location further from the center. 
The chemical profiles of 7.50, 8.00 and 8.25$M_{\odot}$ models at the end of the C-burning phase are shown on Figs. \ref{fig:profiles} and \ref{fig:profiles2}. The core mass are about  1.148, 1.225 and 1.264$M_{\odot}$, respectively. 
The impact on the central $^{20}$Ne and $^{16}$O abundances are  small, near $\sim$ 4\%. 
Higher $^{20}$Ne abundances are associated to lower values of the $^{12}C+^{12}C$ reaction rate due that higher temperatures are needed to ignite carbon. 
Fig. \ref{fig:evol} shows the evolution of several quantities at the position of the flame front.
It can be seen that the flame in the HIN model 
 is characterized by higher temperatures.  
Fig. \ref{fig:hist} show the frequency of $T_{max}$\footnote{Temperature at the base of the flame.} all along the C-burning phase for the three rates adopted, unveiling the similarities in the temperature regimes of C-burning for CF88 and HINRES reaction rates. This explaining the very similar core chemical profiles for $^{20}$Ne and $^{16}$O. However, we note differences in the position of the O/Ne chemical transitions that could lead to different pulsation patterns of ultra-massive WDs. 
Finally Fig. \ref{fig:perfil-branch}, we show the chemical profiles of the 7.50 and 8.00$M_{\odot}$ models for the CF88 reaction rate with branching ratios of 65-35 (filled line) and 56-44 (dashed line). The induced impact of the selected branching ratios on the chemical profiles averages 10\% in the center.  


%
\begin{center}
\includegraphics[width=1\columnwidth]{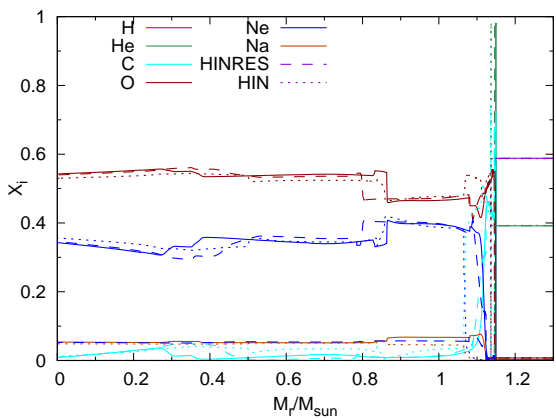}
\captionof{figure}{Chemical profiles for the most abundant species for the 7.50$M_{\odot}$ model. Filled lines correspond to the standard value CF88, while dashed and pointed lines correspond to the HINRES and HIN values from \cite{2022A&A...660A..47M}.}
\label{fig:profiles}
\end{center}

\begin{center}
\includegraphics[width=0.49\columnwidth]{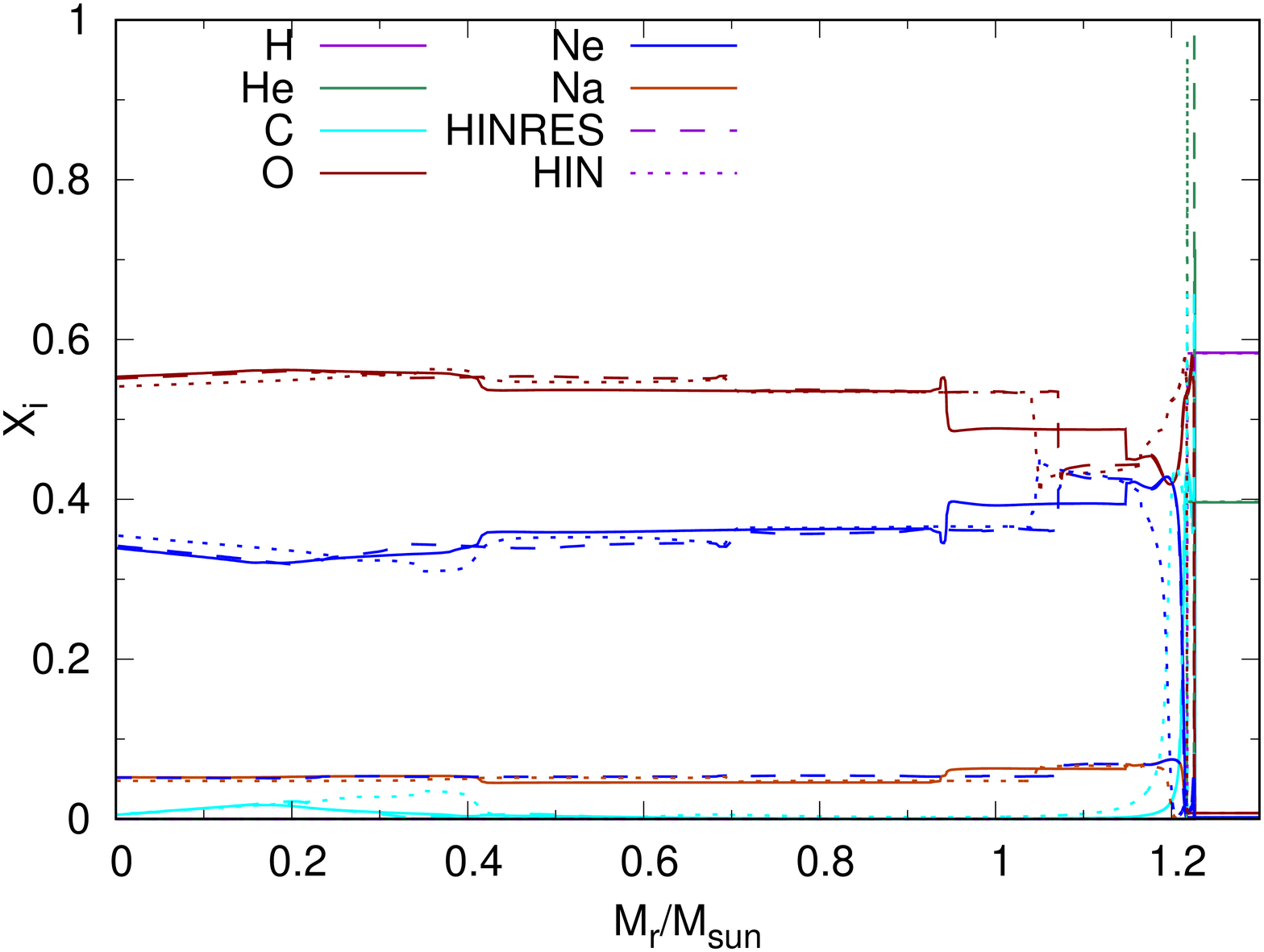}
\includegraphics[width=0.49\columnwidth]{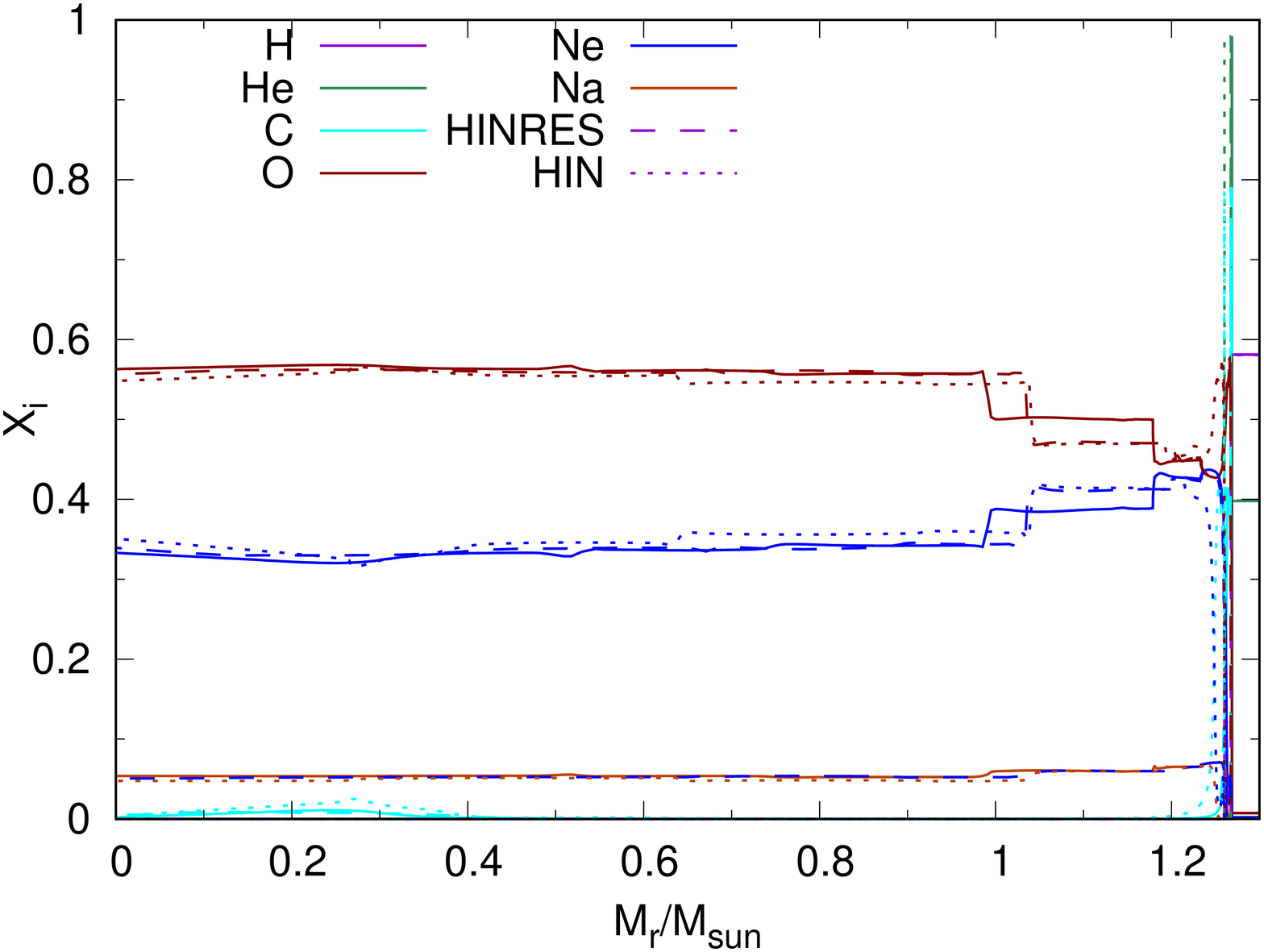}
\captionof{figure}{Same as Fig. \ref{fig:profiles}, but for the 8.00 and 8.25$M_{\odot}$ models, left and right panels respectively.}
\label{fig:profiles2}

\end{center}

\begin{center}
\includegraphics[width=1\columnwidth]{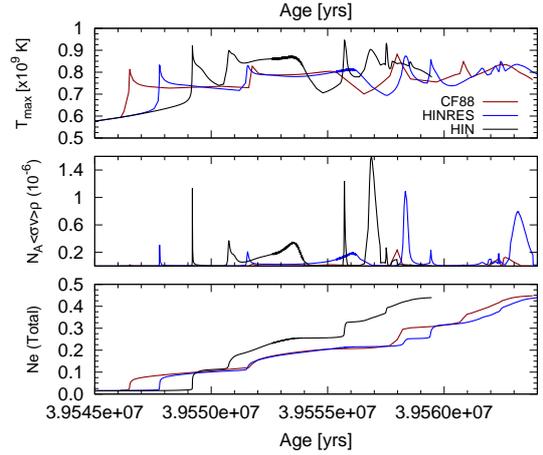}
\captionof{figure}{Evolution of $T_{max}$, $N_A<\sigma v>\rho$ and total $^{20}$Ne content for the 8.00$M_{\odot}$ model for the three nuclear reaction rates adopted.}
\label{fig:evol}

\end{center}

\begin{center}
\includegraphics[width=1\columnwidth]{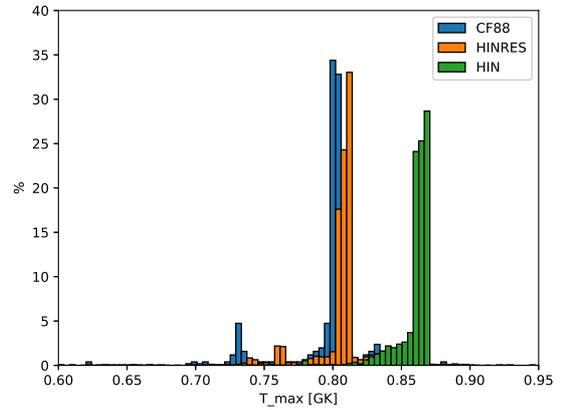}

\captionof{figure}{Histogram of $T_{max}$ for the three nuclear reaction rates adopted of our 8.00$M_{\odot}$ model.}
\label{fig:hist}
\end{center}

\begin{center}

\includegraphics[width=0.49\columnwidth]{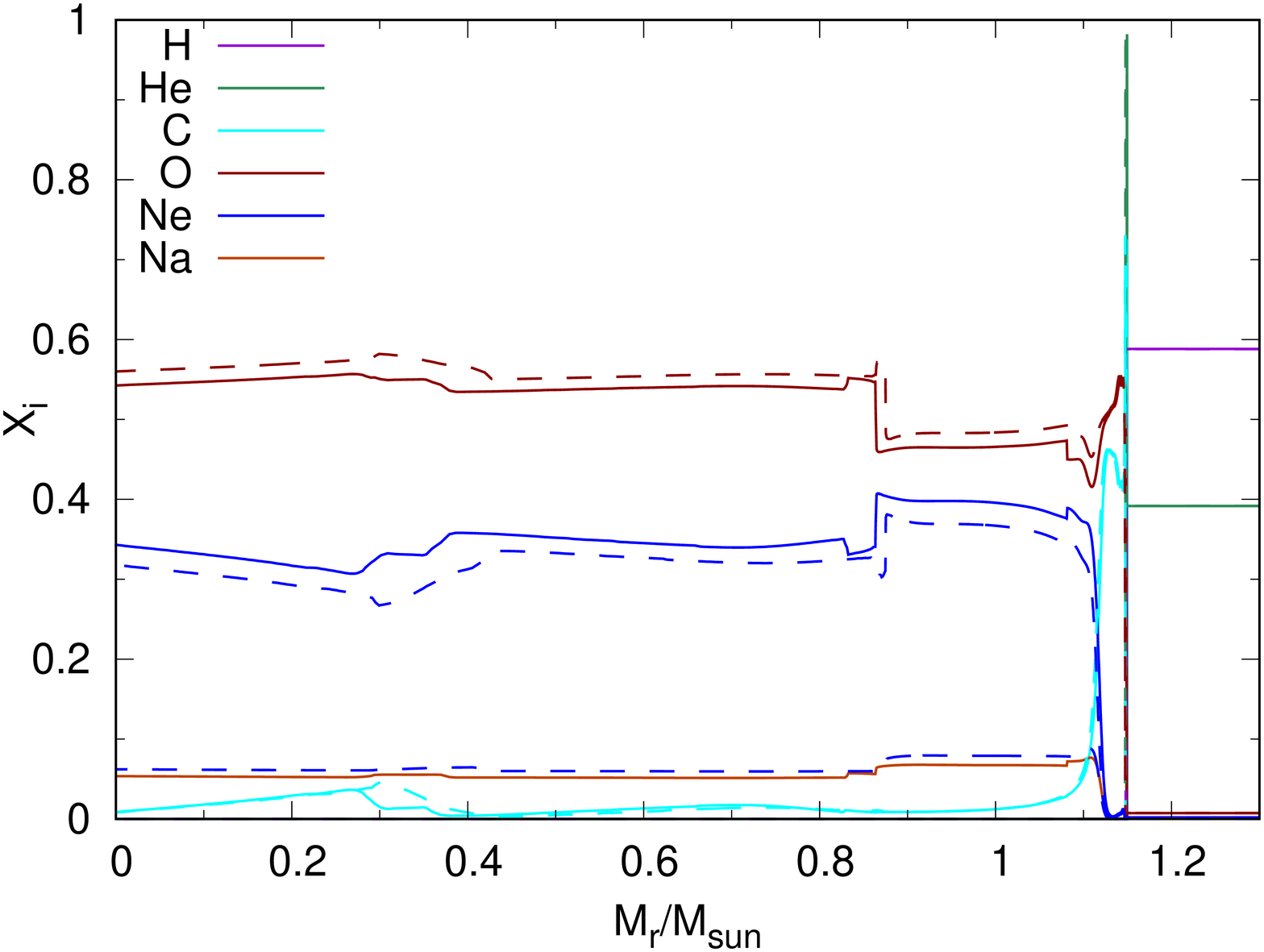}
\includegraphics[width=0.49\columnwidth]{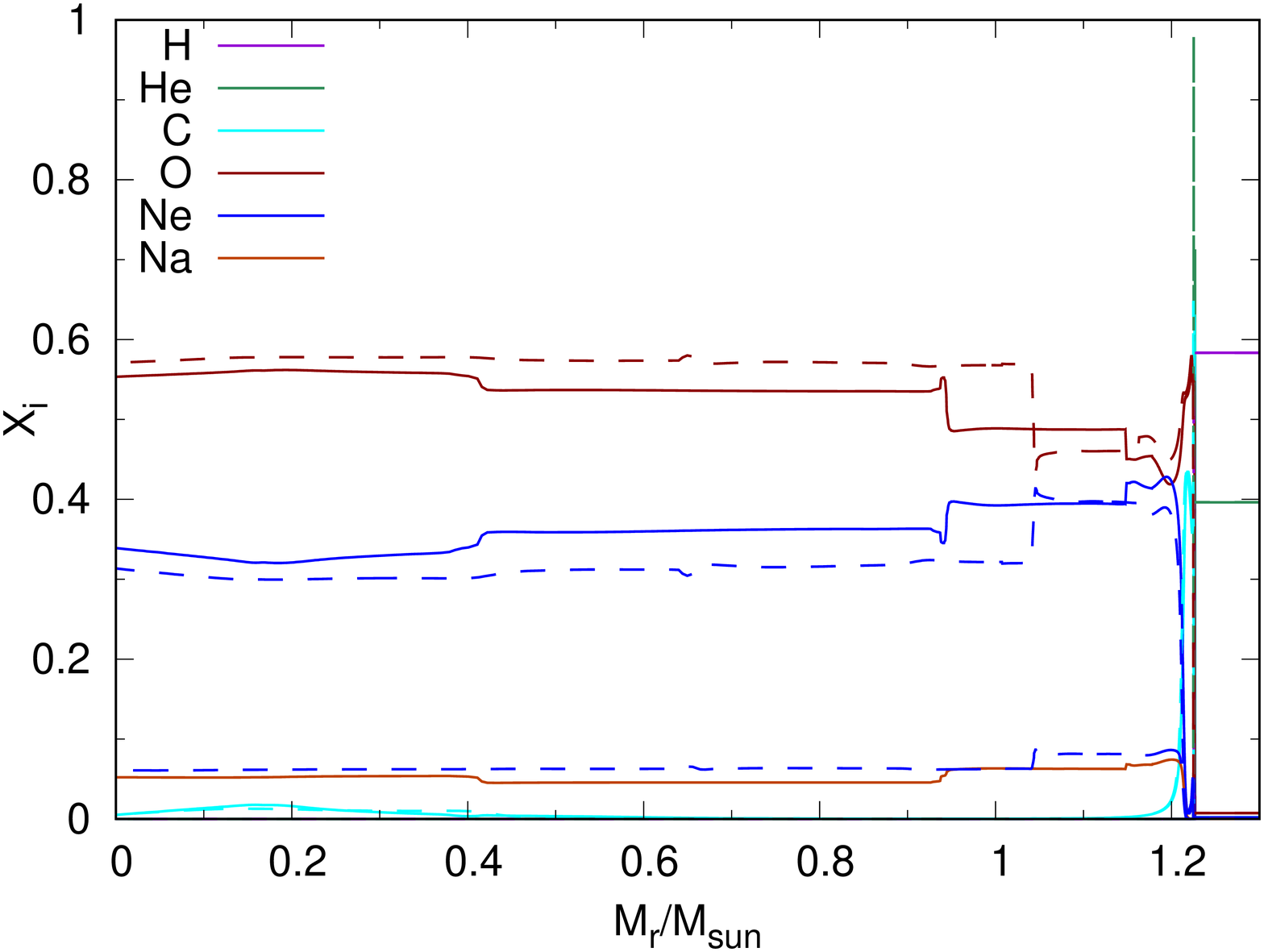}
\captionof{figure}{Chemical profiles for 56-44 (dashed line) and 65-35 (solid line) branching for the 7.50 and 8.00 $M_{\odot}$ models (left and right panel, respectively).}
\label{fig:perfil-branch}
\end{center}

\section*{Conclusions}
We computed the complete evolution of progenitor stars from ZAMS to the end of the C-burning, varying the values of the reaction rates for the $^{12}$C+$^{12}$C nuclear reaction as shown in Fig. \ref{fig:rates}. Our results shows that lower reaction rates leads to a late C-ignition that occurs further out form the center. In consequence, the minimum initial mass for C-burning to occur is shifted $\sim$0.05$M_{\odot}$ for the HIN rate.
Although differences in the most abundant species are small, averaging 4\% in the center,  we note that the location of the O/Ne chemical transition is affected. Additionally, we explored the relevance of the branching ratios for the final chemical structure. We found that differences in the central abundances of $^{16}$O and $^{20}$Ne  remains below 10\%. These features could lead to different pulsation patterns in ultra-massive WDs, that we will explore in the near future. 







\bibliography{degeronimo}
\bibliographystyle{aa}


\end{document}